\documentclass[twocolumn,linenumber]{opjs}
\usepackage{xcolor}
\definecolor{blue4}{rgb}{0.0,0.0,0.54}
\usepackage[colorlinks=true,urlcolor=blue4,linkcolor=blue4,citecolor=blue4]{hyperref}
\usepackage{orcidlink}

\begin{document}

\title{CHARIZARD: A High-throughput Generic Pipeline for Present and Upcoming Interferometers}

\author{Arpan Pal\,\orcidlink{0009-0007-8409-4233}}
\email{arpan5222000@gmail.com}

\author{Ruta Kale\,\orcidlink{0000-0003-1449-3718}}
\email{ruta@ncra.tifr.res.in}

\affiliation{National Centre for Radio Astrophysics, Tata Institute of Fundamental Research, S. P. Pune University Campus, Ganeshkhind, Pune 411007, India}

\begin{abstract}
Radio interferometers now deliver wide bandwidths and high sensitivity, and
with them data volumes that are awkward to reduce by hand and slow to reduce
automatically. The limit is not always the algorithms themselves but the way
the work is laid out across the computing available. We present CHARIZARD, a
pipeline that carries a raw measurement set through flagging,
cross-calibration, self-calibration and imaging to full-Stokes images, and
which is built around distributing that work. The observing band is
repartitioned into sub-bands that stay independent through
cross-calibration, the unit of work becomes the target field and sub-band
pair at self-calibration, and stages touching disjoint data are run
together. Each stage is an ordinary PBS or SLURM job, submitted and tracked
by a scheduler-agnostic layer, and every computation runs inside a single
rootless container, so a user holding no administrative rights gets the same
behaviour on any cluster. We benchmark the pipeline on MeerKAT measurements with full polarization calibration. Residual on-axis
instrumental polarization is $0.13$ per cent in linear and $0.06$ per cent in
circular; the deepest field reaches $4.11\,\mu$Jy~beam$^{-1}$ over $787$~MHz
with $\sigma_I/\sigma_V = 1.010$; positions agree with RACS-low to better
than $0.4\arcsec$ and flux densities to a median ratio of $1.06$. A
$1.5$~TB track completes $48.6$ core-hours of work in $18.4$~hours with self-calibration and polarization calibration. The
next generation of the radio telescopes will not be short of data. CHARIZARD attempts to show how the distribution of the work provides a useful alternative for handling the staggering processing requirements.
\end{abstract}
\section{Introduction}

Radio interferometry measures Fourier components of the sky, i.e. visibilities and then reconstructs the sky brightness by inverting them \citep{thompson2017}. This is a widely used technique to study a variety of cosmic phenomena, starting from the local star, the Sun, to the signatures of the Big Bang. The processing of the radio interferometric data involves steps from correcting the instrumental calibration factors and then inverting the visibilities to the desired image, with special care to not include bad data (RFI) in the processing at every stage. The processing itself is not monotonic and, depending on the quality of the data and the telescope, can require manual inspection and going back and forth in different directions multiple times.

With the recent advancement of radio telescopes such as the uGMRT \citep{gupta2017}, LOFAR \citep{vanharlem2013} and MeerKAT \citep{jonas2016}, we enjoy wide bandwidth, excellent resolution and sensitivity. But with that comes handling of large data volumes and complex instrumental systematics. The advent of this big data era in radio astronomy invites a careful relook at automated processing workflows.

Examples of automated radio imaging pipelines are plenty, and they are typically designed by looking at a specific science or telescope need. This helps to understand the requirements and exploit the telescope-specific advantages. As notable examples, we can mention CAPTURE \citep{kale2021} for the uGMRT, the scripted calibration and imaging pipeline for the VLA \citep{kent2020}, processMeerKAT \citep{collier2021} and CARACal \citep{jozsa2020} for MeerKAT, and the e-MERLIN CASA pipeline \citep{moldon2021}. This list is neither complete nor the best pipeline representation for the mentioned telescopes.

The next generation of radio interferometry demands designing creative solutions to process the huge amounts of data. This problem can be thought of as being addressed in two ways. First, the individual processing components like the software and the file system can be improved on their own. As examples, better file systems which support native parallel read and write can be used instead of the traditional table format for storage. On the software side, as example, to image a large field of view with a large resolution, one might use specific hardware such as Graphical Processing Units (GPUs) to accommodate the requirements or design better calibration algorithms that are fast and accurate for a large number of antennas \citep{smirnov2015}. As examples, \textsc{LibRA} \citep{bhatnagar2025} and \textsc{CubiCal} \citep{kenyon2018} attempt to address some of these issues. One might also come up with creative algorithms such as Image Domain Gridding \citep{vandertol2018}, where it is embarrassingly parallel to take care of the W-terms when the antenna numbers are staggering.

The other way to look at this problem is to come up with different ways to distribute the processing loads to maximize the software performance given the compute available and processing needs. One trivial way to exploit this thinking is to break the huge file size into manageable chunks and treat them independently throughout. One example of this methodology would be processMeerKAT \citep{collier2021}. This pipeline breaks the frequency axis into independent spectral windows and then process them independently, which helps to maintain a high throughput in return. One might also find processing steps where the steps are independent; for example, flagging on the source scans and solving for calibration tables are inherently separate. For a multiple target-source dataset, one can also treat every field individually during self-calibration and imaging, easing the memory load and providing more control for failures.

In this context, we introduce Calibration and Highly Automated Radio Imaging with polariZation and Advanced Resource Distribution (CHARIZARD), as a generic radio interferometric processing pipeline to process present radio interferometers while keeping an eye on the upcoming ones such as the SKA \citep{dewdney2009} and the ngVLA \citep{murphy2018}. CHARIZARD combines smart distribution with accelerated processing steps to achieve as much throughput as possible. The paper is

\section{Individual Components}
CHARIZARD is written in Python and operates with YAML configuration files. The workflow is fixed, while there is flexibility with the choices made, and methods exist for different failure scenarios. As any typical radio imaging pipeline, it operates as an orchestration layer which allocates processes, tracks, report failures and, in case of success, runs the next step of the process. Below we describe, all the modules and the working philosophy of each of them.
\subsection{Orchestrator}
The whole orchestration happens with a module called \textsc{Housekeeper} \footnote{ \url{https://github.com/arpan-52/housekeeper}}. Housekeeper is a cross-platform job orchestration system that relies on existing PBS or SLURM scheduling systems. It offers top-level API methods for both PBS and SLURM, with the additional feature of defining graphs where the user can define dependent processes to wait before a process gets submitted. It can look for specific output products to check if the requirement of the job completion are fulfilled or not. Along with it, it also can look at the job output logs to check if the job has failed or succeeded. User has the flexibility to ignore some errors as well, if, they are confident that it does not hamper the desired outputs. It also provides an interface to check for the standard scheduler-level job operations, where the user can check if the jobs got scheduled and then failed due to resources or some other node-level issues. The code is written to be a pipeline manager, and it dumps all its outputs to a database file which can also be queried anytime. Housekeeper serves as the main connecting and running framework between all the processes in CHARIZARD.
\subsection{Processing Components}
Below we describe all the components which are typical for any radio interferometric image processing.
\subsubsection{Calibration}
\label{sec:calibration}
Radio interferometric calibration involves solving for different instrumental effects by observing suitable calibrators. CHARIZARD offers automated calibration of both total intensity and polarization measurements. It uses the Common Astronomy Software Applications package \citep[CASA;][]{mcmullin2007, casa2022} for solving and applying the calibration.

CHARIZARD supports both linear and circular feed measurements. For total intensity modes, the steps are similar and involves solving for bandpass, residual delays and gains. While for full polarization calibration, for circular feeds, it also involves solving for leakages, cross-hand residual delay and phases, following the polarimetric measurement equation \citep{hamaker1996, sault1996, smirnov2011}. For linear feeds, the same steps are also included, but the order is best suited to match the corruption order.

CHARIZARD resolves the field names from the measurement set and then matches the positions against a publicly available calibrator list \footnote{ \url{https://github.com/arpan-52/VLA-Calibrator-Database-Tools}}, which is predominantly the scraped version of the VLA Calibrator database. Standard flux calibrator models \citep{perley2013, perley2017} are also shipped with the pipeline, and custom models can also be supplied for any calibrator with the help of a different YAML file. CHARIZARD uses suitable uv-ranges for calibration, which are resolved from the same scraped JSON file. User has the final authority to tune the uv-ranges and choice of the calibrators in the configuration files.

\subsubsection{Flagging}
\label{sec:flagging}
For the RFI detection, we use a package called \textsc{Catboss}\footnote{ \url{https://github.com/arpan-52/catboss}}, which has two modules, \textsc{Parallelized Optimized Outlier Hunter (Pooh)} and \textsc{Non-linear Interference Modeling and Korrection Interface (Nimki)}. This package serves as the main workhorse to identify and flag bad samples from the data efficiently.

\textsc{Pooh} looks at the time-frequency amplitude dynamic spectra of each baseline and correlation and flags them independently. \textsc{Nimki} instead compares all baselines against one another, by modelling the visibility amplitude as a function of projected baseline length. The two modules are sensitive to different kinds of corruption, and the pipeline uses both.

\textsc{Pooh} operates on the dynamic spectrum $A(t,\nu)$ of a single baseline and correlation, and proceeds in three stages. The first is bandpass normalisation. The instrumental bandpass and the source spectrum together impose smooth structure across the band. No useful outlier test can be applied until it is removed. We form the time-median spectrum $\bar{A}(\nu) = {\rm med}_t\, A(t,\nu)$, and fit it with a fifth-order polynomial. The fit is iterated: at each iteration the residuals about the current fit are formed, a robust scatter is estimated from their median absolute deviation as $1.4826\,{\rm MAD}$, and channels deviating by more than $5\sigma$ are rejected and excluded from the next iteration. The procedure stops when no new channels are rejected, or after a user-provided number of iterations. Channels rejected in this way are almost always strong persistent carriers, and they are flagged in full at this stage. The dynamic spectrum is then divided by the final fit to give the normalised spectrum $\hat{A}(t,\nu)$.

The second stage sets a detection threshold for each channel,

\begin{equation}
\label{eq:thresh}
\chi_1(\nu) \;=\; \sigma \, {\rm med}_t \, \hat{A}(t,\nu) ,
\end{equation}

where $\sigma$ is 5 or 6 depending on the stage of the pipeline. The threshold is set per channel rather than globally because the noise level is not assumed to be uniform across the band.

The third stage is the detection itself, for which three methods are available. The default is \textit{SumThreshold} \citep{offringa2010, offringa2012}. Interference is frequently not confined to isolated bright samples. A transmitter that is only marginally above the noise but persists over many integrations, or extends over many channels, carries far more total power than a single strong sample and is missed entirely by a per-sample test. SumThreshold therefore tests the mean of a window of $M$ consecutive samples, and lowers the threshold as the window lengthens,

\begin{equation}
\label{eq:sumthreshold}
\chi_M \;=\; \chi_1 \, \rho^{-\log_2 M} , \qquad \rho = 1.5 .
\end{equation}

Windows are applied along the time and frequency axes in turn, for each $M$ in the set specified by the pipeline for that stage, listed in Table~\ref{tab:flagstages}, and flags found in one pass are committed before the next pass reads them. A window is evaluated only if at least $30$ \% of its samples are unflagged, and when a window exceeds $\chi_M$ we flag only those samples within it that individually exceed $\chi_M$, rather than the window as a whole. This prevents flags from spreading into clean data adjacent to a genuine detection. \textit{SumThreshold} allows faint but extended interference to be recovered but can also be prone to overflagging. We do not extend the window beyond $M = 8$ in the pipeline to control overflagging in case of typical fields.

The two remaining methods are per-sample tests. Both are evaluated independently for each channel, using only the unflagged samples of that channel in the normalised spectrum $\hat{A}(t,\nu)$, and both construct their own thresholds directly from the sample distribution rather than starting from $\chi_1$.

\textit{MAD} uses the median absolute deviation. Writing $m(\nu) = {\rm med}_t\,\hat{A}(t,\nu)$ for the channel median, and

\begin{equation}
\label{eq:maddef}
{\rm MAD}(\nu) \;=\; {\rm med}_t \left| \hat{A}(t,\nu) - m(\nu) \right|
\end{equation}

for the median of the absolute deviations about it, a sample is flagged when

\begin{equation}
\label{eq:madflag}
\hat{A}(t,\nu) \;>\; m(\nu) + \sigma \cdot 1.4826 \, {\rm MAD}(\nu) .
\end{equation}

\textit{IQR} uses the quartiles instead. Let $Q_1(\nu)$ and $Q_3(\nu)$ be the 25th and 75th percentiles of the channel, and ${\rm IQR}(\nu) = Q_3(\nu) - Q_1(\nu)$ the interquartile range, that is the width of the interval containing the central half of the samples. A sample is flagged when

\begin{equation}
\label{eq:iqrflag}
\hat{A}(t,\nu) \;<\; Q_1(\nu) - k \, {\rm IQR}(\nu)
\qquad \textrm{or} \qquad
\hat{A}(t,\nu) \;>\; Q_3(\nu) + k \, {\rm IQR}(\nu) ,
\end{equation}

where the multiplier $k$ defaults to $1.5$. This test is two-sided, and so removes low-amplitude excursions as well as high ones.

For efficiency, \textsc{Pooh} processes baselines in batches. All correlations of a batch are passed to the GPU as a single array of shape $(N_{\rm bl}N_{\rm corr}, N_t, N_\nu)$, and the following batch is read from disk while the current one is being processed. The batch size is chosen at run time from the free GPU and host memory, so that the same settings run without modification on a small card. Where no GPU is present the same algorithms are compiled for the CPU through \textsc{Numba} \citep{lam2015}. The two paths are not bit-identical for SumThreshold. The CPU kernel updates the flag array in place as it sweeps, so a window sees flags raised by windows earlier in the same pass, whereas the GPU kernel accumulates candidates in a scratch buffer and merges them only at the end of each pass, which is required to avoid a write race between overlapping threads. The two agree exactly at $M = 1$ and diverge slowly thereafter; over a representative batch with $M \in \{1,2,4,8\}$ they disagree on $<1.0$ \% of samples.

The products on which \textsc{Pooh} operates need not be those recorded in the measurement set. It can also form two derived, or virtual, correlations and run the same detection on them. These are Stokes $V$, which denotes the circular polarization, and a linear polarization product $P$, formed from the cross hands as $(XY+YX)/2$ for linear feeds and $(RL+LR)/2$ for circular feeds.

A virtual sample is treated as valid only where every constituent correlation is unflagged, and a flag raised on a virtual product is written back to its constituent correlations, or optionally to all four. The motivation is that astronomical sources are very rarely circularly polarised at more than the per cent level, whereas terrestrial transmitters frequently are. In $V$ the source largely cancels while the interference does not, so the interference becomes the dominant signal in the plane rather than a small perturbation on top of a bright source. $P$ performs the same role for linearly polarised interference.

\textsc{Nimki} looks at the `uv' domain for the RFI flagging. \textsc{Nimki} works on time chunks of $30$ minutes by default. Within each chunk, and separately for each spectral window and correlation, it collects every unflagged amplitude together with its projected baseline length $r = \sqrt{u^2+v^2}$ in wavelengths. Because $r$ is evaluated per channel, all channels of the chunk contribute to a single distribution of amplitude against $r$. To this distribution, it fits

\begin{equation}
\label{eq:gabor}
\tilde{V}(r) \;=\; \sum_{i=1}^{n} A_i \, \exp \left( -\frac{r^2}{2\sigma_i^2} \right) \cos ( \omega_i r + \phi_i ) ,
\end{equation}

a sum of Gabor functions, with $n = 5$ by default. The fit is performed by Levenberg--Marquardt with two random restarts to reduce sensitivity to local minima. An adaptive mode is also available, in which components are added one at a time until the fractional improvement in the residual falls below $5$ \%. A Gabor basis is used because the amplitude of a resolved source oscillates with baseline length.

Outliers are then identified by a robust clip on the residual,

\begin{equation}
\label{eq:mad}
\left| (A - \tilde{V}) - {\rm med} \right| \;>\; \sigma \cdot 1.4826 \, {\rm MAD} ,
\end{equation}

The fit and the clip are implemented in C++, and chunks are processed in parallel across cores. \textsc{Nimki} is currently CPU-only by design.

\begin{deluxetable}{llcc}
\tablecaption{Where the flaggers run and with what settings. \label{tab:flagstages}}
\tablehead{
\colhead{Stage} & \colhead{Flagger} & \colhead{$\sigma$} & \colhead{$M$}
}
\startdata
Before calibration, calibrators & Pooh & 6 & $\{1,2\}$ \\
Before calibration, targets & Pooh & 6 & $\{1,2\}$ \\
After round 1, calibrators & Pooh + Nimki & 5 & $\{1,2,4,8\}$ \\
Final, all fields & Pooh + Nimki & 6\tablenotemark{a} & $\{1,2\}$ \\
Each self-calibration round & Pooh & 5 & $\{1,2,4,8\}$ \\
\enddata
\tablenotetext{a}{\textsc{Pooh} runs at $\sigma = 6$ in this pass; \textsc{Nimki} runs at $\sigma = 5$.}
\end{deluxetable}

Flagging is the longest single stage of a normal run, so its throughput is worth quoting. We measured it on a MeerKAT sub-band of $61\,161$ rows $\times$ $432$ channels $\times$ 4 correlations, that is $2.64 \times 10^{7}$ visibilities, or $1.057 \times 10^{8}$ visibility samples counting each correlation together. Both modules were driven through the production command-line interface on a fresh copy of the measurement set, so the timings include reading the data, forming the products, detecting, and writing the \texttt{FLAG} column back, which is what the pipeline actually pays. On a single workstation with an Intel Core Ultra 9 185H and an RTX 4060 laptop GPU, \textsc{Pooh} took $17.6$ s in GPU mode and $17.0$ s in CPU mode, corresponding to $6.0$ and $6.2$ million samples per second, or about $1.5$ million visibilities per second. \textsc{Nimki} took $37.2$ s on the same data, or $2.8$ million samples per second.

To separate the cost of detection from the cost of data handling, we also timed the kernels alone, on synthetic dynamic spectra held in memory, with compilation excluded and, in the GPU case, with the arrays already resident on the device. Table~\ref{tab:kernels} gives the result. SumThreshold sustains $6.7 \times 10^{7}$ visibilities per second on the CPU and $6.4 \times 10^{7}$ on the GPU. The two per-sample tests are faster by between one and two orders of magnitude, reaching $2.3 \times 10^{9}$ and $1.8 \times 10^{9}$ visibilities per second on the CPU and $3.7 \times 10^{9}$ and $2.8 \times 10^{9}$ on the GPU, since each performs a single comparison per sample against a threshold that has already been computed. The gap is the price of the window accumulation of Equation~\ref{eq:sumthreshold}, which sweeps the data twice for every value of $M$ and is accordingly some forty times more expensive than a per-sample test. It buys the sensitivity to faint but extended interference described above, and on any realistic dataset it is paid out of I/O time rather than added to it.

\begin{deluxetable}{lccc}
\tablecaption{Detection kernel throughput, measured in isolation from measurement set I/O on synthetic dynamic spectra of $4.0 \times 10^{7}$ samples ($64$ integrations $\times$ $432$ channels per baseline and correlation), best of nine passes with compilation excluded and, for the GPU, with the data already resident on the device. Rates are quoted in units of $10^{6}$ visibilities per second, taking four correlations to the visibility. \label{tab:kernels}}
\tablehead{
\colhead{Method} & \multicolumn{2}{c}{detection kernel} & \colhead{statistics} \\
\colhead{} & \colhead{CPU} & \colhead{GPU} & \colhead{(CPU)}
}
\startdata
SumThreshold &   67 &   64 &  35 \\
MAD          & 2320 & 3720 &  71 \\
IQR          & 1770 & 2820 & 120 \\
\enddata
\end{deluxetable}

The GPU is of little help to SumThreshold. At $M \le 8$ the windows are short and the memory access is strided, so the kernel is limited by bandwidth rather than by arithmetic and the device offers no advantage over the 22 CPU threads. It gives about a factor of $1.6$ on the two per-sample tests, which are already fast enough for the difference not to matter. This is consistent with the end-to-end timings, where the GPU and CPU modes were also indistinguishable.

Comparing the two sets of measurements identifies where the time is spent. Detecting the whole MeerKAT sub-band requires $1.057 \times 10^{8}$ samples to pass through the SumThreshold kernel, which at the measured rate takes $0.40$ s on the CPU and $0.42$ s on the GPU, against end-to-end times of $17.0$ and $17.6$ s. The detection kernel therefore accounts for about $2$ \% of what the pipeline actually spends, and the per-channel statistics for a further $4$ \%. The remaining $93$ \% is reading the measurement set, forming the products, and writing the \texttt{FLAG} column back. A second measurement confirms that this residue is dominated by I/O: repeating the run with identical settings on a $15$ GB measurement set, $4.6$ times larger, the rate fell to about $1.0$ million samples per second, so that the job took $27$ times longer for $4.6$ times the data. Once the working set no longer fits comfortably in memory, throughput is set by the disk. The practical consequence is that optimising the detection kernels further would yield almost nothing, whereas keeping each job small enough to remain in memory yields a great deal.

\subsubsection{Imaging}
\label{sec:imaging}

For imaging and deconvolution CHARIZARD uses \textsc{WSClean} \citep{offringa2014}, which implements \textsc{Clean} \citep{hogbom1974} over a w-stacking gridder. We chose it because the W-stacking algorithm is considerably faster. The pipeline calls \textsc{WSClean} in three different situations, starting with making a dirty image, then getting the model for self-calibration and then for the final imaging.

The settings we use are the same in all three cases apart from the depth. We use Briggs robust 0.0 \citep{briggs1995} weights for the visibilities, which is our compromise between sensitivity and resolution and works reasonably for most fields. It is still not configurable but is a priority for the future releases. The image size and the cell size are set by the user in the configuration file.

We always use 4 output channels with \texttt{-channels-out 4} and clean them jointly with \texttt{-join-channels}. Further, we fit a third-order polynomial in frequency to each clean component with \texttt{-fit-spectral-pol 3}. This keeps the spectral behaviour of the components physical and stops the individual channel images from becoming noisy at the edges of the band.

Extended emission is handled with multiscale cleaning \citep{cornwell2008, offringa2017} through \texttt{-multiscale}, with the scale bias at 0.6. We let \textsc{WSClean} pick the scales itself. The loop gain is 0.1 and we use \texttt{-mgain 0.7}, so that each major cycle takes out 70 \% of the peak before going back to the visibilities. We use the automatic masking of \textsc{WSClean}, with \texttt{-auto-mask 5} and \texttt{-auto-threshold 3}. We also pass an explicit threshold from the self-calibration schedule, but since \texttt{-auto-threshold 3} is always active, the effective stopping point is the shallower of the two, so the schedule can only ever be more conservative than the noise. In addition to the automatic mask the user can supply a FITS mask, and CHARIZARD can build one from the TGSS \citep{intema2017} and NVSS \citep{condon1998} catalogues through a helper module called \textsc{Foresight} \footnote{ \url{https://github.com/arpan-52/foresight/}}.

At the end of the self-calibration loop, in the case of successful polarization calibration, the final imaging is repeated for Stokes $I$, $Q$, $U$ and $V$ separately. The Stokes $I$ run is also asked for a source list with \texttt{-save-source-list}.

\section{Pipeline Workflow}
\label{sec:workflow}

\begin{figure*}
\centering
\includegraphics[width=\textwidth]{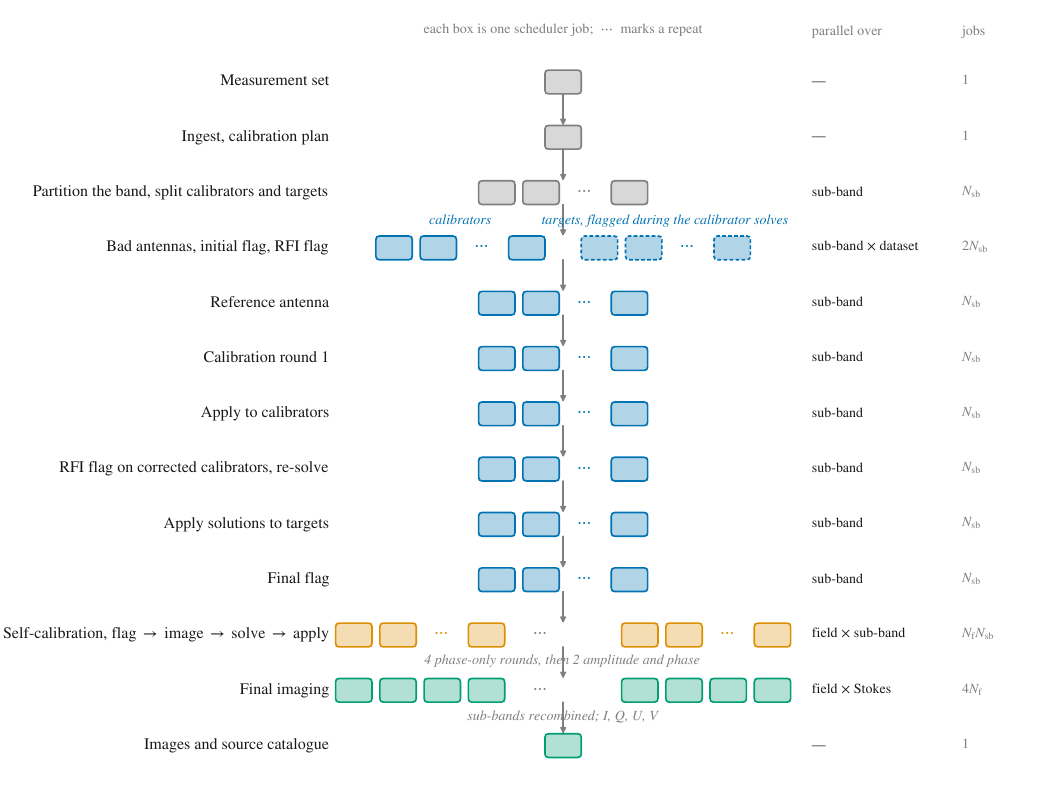}
\caption{The CHARIZARD workflow. Each box is one scheduler job and an
ellipsis marks a repeat, so the width of a row is the number of jobs that
stage runs at once; the axis that repeat runs over is named on the right.
Reading that column downwards gives the design of the pipeline. The dashed group
is the one point where two disjoint datasets are in processing together. At self-calibration the unit of work
changes, since separate target fields share no solution, and again at the
final imaging, where the sub-bands are recombined and the four Stokes
parameters are imaged independently. $N_{\rm sb}$ is the number of sub-bands
and $N_{\rm f}$ the number of target fields.}
\label{fig:flow}
\end{figure*}

The preceding section described the components in isolation. We now set out the order in which a run executes them (Figure~\ref{fig:flow}). Configuration selects which optional stages are enabled and supplies the numerical parameters each one receives, but it does not expose the sequence.

\subsection{Ingest and calibration planning}
\label{sec:ingest}

A run begins with a metadata pass over the measurement set which records the fields, spectral windows, channel counts, antennas and correlation products. The correlation types identify the feed basis, circular or linear. Field names are then compared against the standard flux calibrators, and the remaining fields are matched by position against the calibrator database described earlier. A positional match designates a phase calibrator and carries with it the recommended $uv$ range for that calibrator in the observing band, which is applied to every solve involving that field. Fields without a match are treated as science targets. Full polarization calibration is enabled only when a leakage calibrator, a polarization angle calibrator, and a full Stokes model for the latter are all present; if any of the three is missing, the pipeline records the reason and continues in total intensity mode.

The resulting assignments are written to a YAML file alongside the data. Subsequent stages read this file rather than repeating the analysis, so that self-calibration obtains the surviving correlation count and the chosen reference antenna without re-deriving them. The file also serves as a machine-readable record of every decision the pipeline made.

\subsection{Frequency partitioning}
\label{sec:chunking}

The partitioning of the band determines the throughput of the entire run. Spectral windows as delivered by the correlator reflect the instrument configuration. Five \% of the channels at each spectral window edge are removed, since the response rolls off there and those channels are seldom usable. The remainder is treated as a single contiguous block and divided into $N_{\rm sb}$ equal sub-bands, with $N_{\rm sb}$ specified by the user. Each sub-band is written as an independent measurement set by \texttt{mstransform}, and the splits execute concurrently, one job per sub-band.

Every sub-band is written twice: once containing the calibrator fields and once containing the targets. The additional disk cost is modest, and the benefit is structural, since the calibrator and target visibilities thereafter reside in separate files. This separation is what permits the target fields to be flagged while the calibrator solutions are being computed, as the two jobs address disjoint data.

A further reduction is available when polarization calibration is not being performed. The cross-hand correlations are then of no use downstream, and the split retains only the parallel hands, $RR,LL$ or $XX,YY$, halving the volume handled by every subsequent job.

The choice of $N_{\rm sb}$ involves a genuine trade off. Increasing it raises the number of concurrent jobs and lowers the memory required by each, but every sub-band must remain wide enough for the bandpass and leakage solutions within it to retain adequate signal to noise, and narrow enough that those solutions vary smoothly across it. For the uGMRT and MeerKAT datasets processed here we select $N_{\rm sb}$ such that each sub-band spans a few tens of MHz. The concurrency the scheduler will actually grant provides the other bound: once $N_{\rm sb}$ exceeds the number of simultaneously executing jobs, further subdivision yields no additional speed.

After the split the sub-bands are independent throughout cross-calibration. No bandpass, delay, gain or leakage solution in one sub-band depends on any other. Each pipeline stage is therefore a set of $N_{\rm sb}$ jobs submitted together, and the elapsed time of the stage approaches that of a single sub-band whenever the scheduler admits that many jobs at once.

\begin{table*}
\centering
\caption{The three benchmark observations.}
\label{tab:obs}
\begin{tabular}{lccc}
\hline
\hline
Date & 2026 Feb 28 & 2026 Mar 1 & 2026 Jun 7 \\
Filter sub-band & S0 & S0 & S1 \\
Frequency (MHz) & $1750$--$2625$ & $1750$--$2625$ & $1968.8$--$2843.5$ \\
Antennas & 58 & 58 & 60 \\
Reference antenna & m033 & m033 & m040 \\
Target fields & J0159.8$-$0849 & J0011.7$-$1523 & ACT-CL~J0102$-$4915$^{a}$ \\
 & J0242.5$-$2132 & J0014.3$-$3022 & \\
Flux density, bandpass, leakage & J0408$-$6545 & J0408$-$6545 & J0408$-$6545 \\
Phase & J0219+0120, & J0025$-$2602 & J0133$-$5200 \\
 & J0242$-$2132 & & \\
Polarization angle & J0521+1638 & J0521+1638 & J0521+1638 \\
Image size, cell & $6144^{2}$, $0\farcs9$ & $9600^{2}$, $0\farcs9$ & $11400^{2}$, $0\farcs8$ \\
Scheduler jobs & 168 & 164 & 130 \\
\hline
\end{tabular}

\par\smallskip
\parbox{\textwidth}{\footnotesize
$^{a}$ The merging cluster El Gordo \citep{menanteau2012}.
}
\end{table*}

\subsection{Cross-calibration}
\label{sec:crosscal}

Processing opens with the non-working antenna search and the initial flag application of Section \ref{sec:flagging}, followed by a conservative \textsc{Pooh} pass over the calibrators.

The reference antenna is selected only after this first excision pass. The ordering is deliberate: chosen on unflagged data, the metric will frequently favour an antenna whose apparent stability is an artefact of the interference it is sitting in. For each antenna we collect $A_a$, the set of median parallel-hand visibility
amplitudes of every baseline involving antenna $a$, and evaluate

\begin{equation}
\label{eq:refant}
S_a \;=\; \frac{{\rm med}(A_a)}{\sigma(A_a)} \left( 1 - f^{\rm flag}_a \right) ,
\end{equation}

where $f^{\rm flag}_a$ is the fraction of that antenna's samples already flagged. The metric is the amplitude stability of an antenna weighted by how much of its data survives. Antennas with fewer than a hundred surviving samples are not scored at all. Each sub-band nominates a candidate, and the majority selection becomes the reference antenna for the run. A reference antenna named in the configuration file overrides the vote.

The first calibration round then proceeds along the chain of Section \ref{sec:calibration}. The target field flagging is submitted concurrently with these solves, since the targets occupy separate files and no calibrator solution touches them; the longest flagging job of the run is thereby hidden behind the calibration and costs no additional elapsed time.

Once the first round completes, its solutions are applied to the calibrators and the corrected calibrator visibilities are flagged a second time, now at the tighter settings and with \textsc{Nimki} included. This second excision is the more productive of the two. Low level interference which was concealed beneath the bandpass shape and the gain variation in the raw data becomes conspicuous once both have been divided out. The full chain is then re-solved against the cleaned data, and only these second-round tables are transferred to the science targets. A final flagging pass is applied to the corrected targets before imaging begins.

\subsection{Self-calibration, imaging and direction dependent calibration}
\label{sec:sc_dd}

Self-calibration \citep{cornwell1981, pearson1984} follows. At this point the unit of work changes. Until this point it was the sub-band; from here it becomes the (target field, sub-band) pair, since separate targets share no solutions. Each pair is extracted into its own measurement set and averaged in frequency by a configurable factor, ten by default.

Each round performs the same four operations in sequence: flag, image, solve, apply. Progression across rounds is governed by a single factor $f$, with a default of two. At each round the clean iteration budget is multiplied by $f$ while the clean threshold and the solution interval are divided by it, both subject to floors which prevent the schedule from running away. The coupling is intentional, since the three quantities are not independently justifiable: a deeper clean is warranted only if the gain solutions have improved, and a shorter solution interval only if the sky model has improved. A single parameter therefore moves all three together, with $f = 1.5$ giving a cautious schedule and $f = 2$ the standard one. The default is four phase-only rounds followed by two amplitude and phase rounds.

Where it is enabled, the direction-dependent stage \citep[see e.g.][]{tasse2018} follows. The sub-bands of a field are recombined, and the final Stokes $I$ image is catalogued with \textsc{PyBDSF} \citep{mohan2015}. Catalogue entries above a configurable flux threshold become the peeling directions. \textsc{Crystalball}\footnote{ \url{https://github.com/caracal-pipeline/crystalball}} predicts the model for each direction into a separate visibility column, after which \textsc{QuartiCal} \citep{kenyon2025} solves a direction-independent delay term together with a direction-dependent gain for every direction in a single pass over the data. The corrupted models of the peeling directions are subtracted, and a final deep \textsc{WSClean} run images the residual column.

\subsection{Failure handling}
\label{sec:failure}

A run comprising several hundred scheduler jobs will encounter failures, and the pipeline requires a defensible policy for them. \textsc{Housekeeper} does not rely on exit codes, since CASA based tasks routinely report errors in their logs while returning success. Each job log is instead parsed in full and compared against a curated list of benign messages, among them leap second warnings and Numba deprecation notices; any error line absent from that list marks the job as failed irrespective of its exit status. Each stage additionally verifies that the products it was expected to create are present before it is recorded as complete. An excessively strict criterion costs a repeated job, whereas an excessively permissive one costs a corrupted final image, and we have resolved the asymmetry in favour of strictness.

The treatment of a failed sub-band is left to the user through a single switch, which we term `brotherhood'. When enabled, which is the default, a stage succeeds only if it succeeded for every sub-band, and any failure halts the run at that point with the offending logs identified. When disabled, failed sub-bands are removed from the active set and processing continues with the remainder.

\section{Results}
\label{sec:results}

We benchmark CHARIZARD on already available MeerKAT observations. We choose S band due to immediate availability and as a general case. The MeerKAT S-band system \citep{kramer2016} covers $1750$--$3500$~MHz, from which an $875$~MHz digital filter selects one of five overlapping sub-bands \citep{ranchod2025}. Two of our observations were recorded in S0 ($1750$--$2625$~MHz) and one in S1 ($1968.75$--$2843.75$~MHz). MeerKAT S band operates with linear feed and data were processed with polarization calibration enabled. We have used four subbands during processing. All the jobs were run on the NCRA \textit{bhima} cluster under OpenPBS, with each job given one node and eight cores. \textit{Bhima} has a total of 4 compute nodes, each with 64 CPU cores in dual-socket configuration and 2~TB RAM (256 cores, 8~TB RAM total). Examplary benchmarking observation details can be found in Table \ref{tab:obs}.

\subsection{Calibration accuracy}
\label{sec:res_cal}
We report the calibration accuracy on the calibrator visibilities. For every 2 MHz of frequency span of every sub-band we vector-average \texttt{CORRECTED\_DATA} over all baselines and times of a calibrator and form $I = (XX+YY)/2$, $Q = (XX-YY)/2$, $U = (XY+YX)/2$ and $V = -i(XY-YX)/2$. On J0408$-$6545, which is an unpolarized calibrator, the residual linear polarization is $0.13 \pm 0.05$ \% and the residual circular polarization $0.05 \pm 0.03$ \%. The worst $2$~MHz bin reaches $0.641$ \%. The recovered
Stokes $I$ sits at $1.00 \pm 0.05$ times the model. On J0521+1638 the
recovered position angle is $-12.54 \pm 0.62 ^{\circ}$  against a model value of
$-12.13^{\circ}$, and the spurious circular polarization is $0.04$ \%. No polarization solution of any kind
is derived on the phase calibrator J0133$-$5200, and it returns
$|V|/I = 0.06$ \%.

\begin{figure*}
\centering
\includegraphics[width=\textwidth]{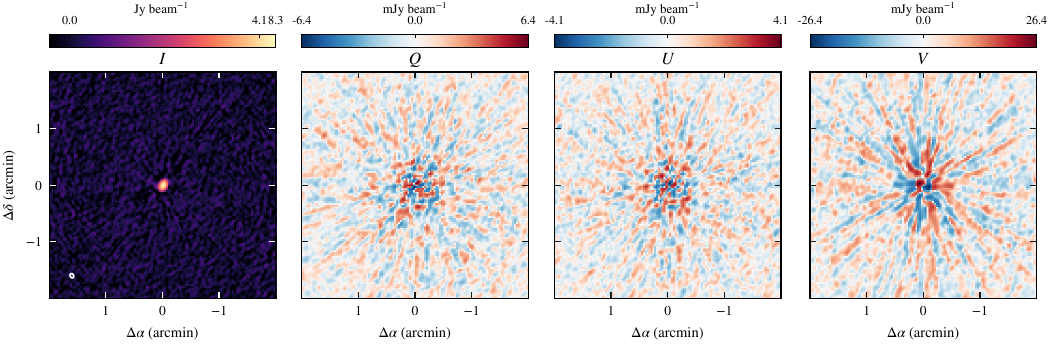}
\caption{The leakage calibrator J0408$-$6545 imaged back from
\texttt{CORRECTED\_DATA} in all four Stokes parameters, over a four
arcminute cutout. The solve treats this source as unpolarized, so the
expected result is a point source in $I$ and nothing in $Q$, $U$ or $V$.
Each panel has its own colour bar and its own asinh scale stretched to its
own peak. The radial striping is the dynamic range limit around an $8$~Jy point source.}
\label{fig:rescalimg}
\end{figure*}

\begin{figure*}
\centering
\includegraphics[width=\textwidth]{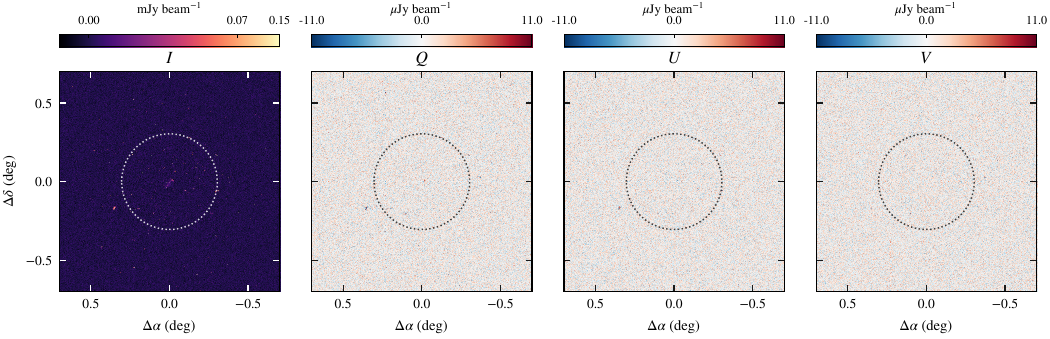}
\caption{The inner $1\fdg4$ of the El Gordo field in Stokes $I$, $Q$, $U$ and
$V$. $Q$, $U$ and $V$ share one symmetric colour scale of five times the
noise; $I$ has its
own asinh scale, clipped well below its peak so that faint sources stay
visible at page size. The dotted circle is the half-power point of the
primary beam..}
\label{fig:resfield}
\end{figure*}

\begin{figure*}
\centering
\includegraphics[width=\textwidth]{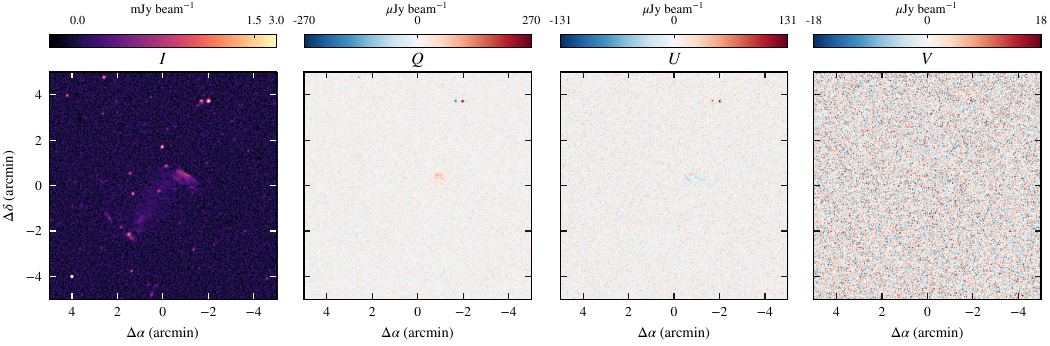}
\caption{The central ten arcminutes, each panel on its own colour scale
stretched to its own peak. $Q$ and $U$ are three orders of magnitude below
$I$, so this scaling is what makes the polarized emission visible.}
\label{fig:reszoom}
\end{figure*}

Imaging the leakage calibrator back from the same corrected data gives the
image-plane equivalent, shown in Figure~\ref{fig:rescalimg}. The calibrator carries $8.26$~Jy at the phase centre, so a leakage of one part in ten thousand can still be seen in the images. Read at the Stokes $I$ peak pixel, the
multi-frequency synthesis images give $Q = -0.47$, $U = -0.77$ and
$V = -1.09$~mJy~beam$^{-1}$ on $8263.7$~mJy~beam$^{-1}$, so $p = 0.011$ and
$|V|/I = 0.013$ \%. Across the sixteen output channels $p$ has a median of
$0.085$ \% and a worst channel of $0.127$ \%, and $|V|/I$ a
median of $0.024$ \% and a worst channel of $0.052$ \%.

All of these figures are on-axis. On the forty brightest sources
of the El Gordo field the median fractional polarization is $1.01$ \%
inside the half-power point of the primary beam and $3.54$ \% outside
it, against $0.13$ \% on-axis. Part of the inner figure is intrinsic to
the sources and it is therefore an upper limit, but the growth with radius is
not. Polarimetry away from the pointing centre needs a beam model that the
pipeline does not currently apply.

\subsection{Image quality}
\label{sec:res_image}

The El Gordo field was imaged with a bandwidth of $787$~MHz with \textsc{WSClean} at the
settings of Section~\ref{sec:imaging}, giving an $11400^{2}$ pixel image at
$0\farcs8$ per pixel, $2\fdg53$ on a side, with a $4\farcs25 \times 3\farcs75$
restoring beam at a reference frequency of $2.406$~GHz. The same
visibilities, weighting and deconvolution produced all four Stokes
parameters. Figure~\ref{fig:resfield} shows the inner $1\fdg4$ and
Figure~\ref{fig:reszoom} the central ten arcminutes.

The noise is $4.11$, $4.10$, $4.07$ and $4.07\,\mu$Jy~beam$^{-1}$ in $I$,
$Q$, $U$ and $V$, measured as $1.4826$ times the median absolute deviation of
the full map. Rather than compare against an assumed system temperature we
use Stokes $V$ as the reference: the sky is circularly polarized far below
this level, so a $V$ image made from the same visibilities with the same
weighting and deconvolution is a probe of the thermal noise directly. The ratio $\sigma_I/\sigma_V$ then gives
the degradation of the total intensity image by everything that is not
thermal noise, such as residual calibration error, deconvolution error,
confusion and the sidelobes of imperfectly subtracted sources. We measure
$1.010$, and $\sigma_Q/\sigma_I = 0.998$. The total intensity image is
within 1 \% of its own thermal floor.

\subsection{Astrometry and flux density scale}
\label{sec:res_astro}

The tests so far compare the pipeline against the models it was given. Matching with an
external catalogue confirms whether the positions and flux
densities agree with an independent instrument. We use the Rapid ASKAP
Continuum Survey \citep{mcconnell2020}, carried out with ASKAP
\citep{hotan2021}, specifically the RACS-low first data
release source catalogue \citep{hale2021}, at $887.5$~MHz with a $25\arcsec$
restoring beam and a median positional uncertainty near one arcsecond in
these fields. The comparison uses the 2026 February 28 observation, whose two
target fields lie well inside the RACS footprint.

We correct the primary beam measured holographically by \citet{devilliers2023}. After primary beam correction, the \textsc{PyBDSF} \citep{mohan2015} was used to make a source catalogue. Cataloguing the corrected images at a $5\sigma$ island and pixel threshold
returns $720$ sources in the two target fields, $407$ in J0159.8$-$0849 and $313$ in J0242.5$-$2132.

Matching is done two ways. For astrometry
we take each catalogued source and find the nearest RACS entry within
$10\arcsec$. For flux density we work outward from RACS.
\begin{figure*}
\centering
\includegraphics[width=\textwidth]{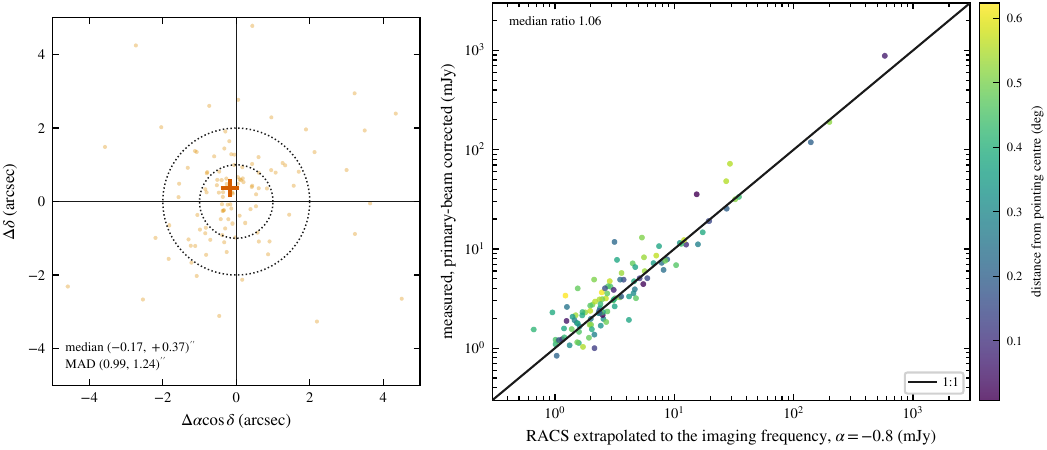}
\caption{The 2026 February 28 images against RACS-low. Left: position
offsets, RACS minus CHARIZARD, for every catalogued source with a RACS
counterpart within $10\arcsec$; dotted circles at $1\arcsec$ and $2\arcsec$,
cross at the median. Right: measured flux density at $2.187$~GHz after the
primary beam correction, against the RACS value
extrapolated to the same frequency with $\alpha = -0.8$, coloured by distance
from the pointing centre. The solid line is equality.}
\label{fig:resastro}
\end{figure*}

\begin{figure*}
\centering
\includegraphics[width=\textwidth]{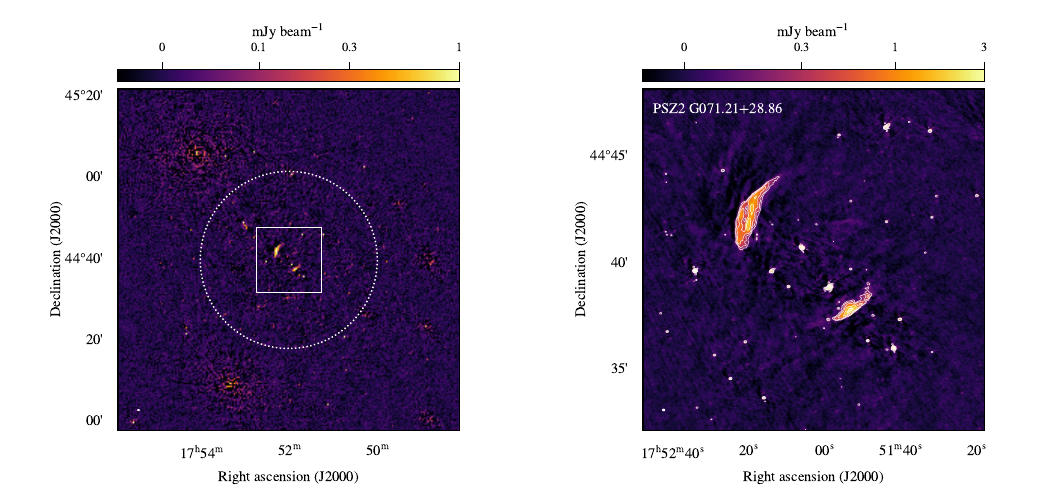}
\caption{The uGMRT band-4 image of the PSZ2~G071.21+28.86 field at
$650$~MHz, before primary beam correction. Left: the inner $1.4^{\circ}$; the
dotted circle is the half-power point of the primary beam and the box marks
the right panel. Right: the central $16$ arcminutes around the cluster, with
contours at $0.165\times2^{n}$~mJy~beam$^{-1}$, starting at five times the
local noise. Both panels use an asinh scale, and the restoring beam is drawn
at the bottom left}
\label{fig:resgmrt}
\end{figure*}
\begin{figure*}
\centering
\includegraphics[width=\textwidth]{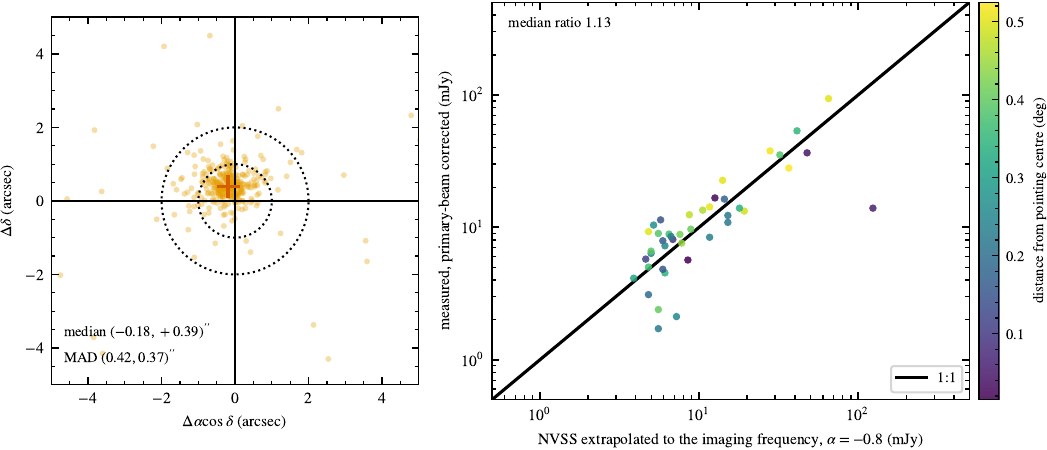}
\caption{The uGMRT image against external catalogues. Left: position
offsets, LoTSS-DR2 minus CHARIZARD; dotted circles at $1\arcsec$ and
$2\arcsec$, cross at the median. Right: measured flux density at $650$~MHz
after primary beam correction, summed within $30\arcsec$ of each NVSS
position, against the NVSS value extrapolated with $\alpha = -0.8$, coloured
by distance from the pointing centre. The point far below the line is the
north-eastern arc of the cluster, which NVSS lists as a single component
$2\arcmin$ across. The solid line is equality.}
\label{fig:resgmrtastro}
\end{figure*}

Of the $720$ catalogued sources, $113$ have a RACS counterpart within
$10\arcsec$. The median offset is $-0.17\arcsec$ in right ascension and
$+0.37\arcsec$ in declination. Against a four to six arcsecond beam this is
under a tenth of a beam, and it is comparable to the positional uncertainty
of RACS itself. The astrometry is limited by the reference rather than by
the pipeline.

The flux comparison needs an assumed spectral index, since RACS-low carries
no spectral information and $887.5$~MHz to $2.187$~GHz is a factor of $2.5$.
At $\alpha = -0.8$ the median ratio of measured to extrapolated flux density
over the $112$ matched entries is $1.06$, with the two fields at $1.08$ and
$0.99$. The ratio shows no trend with radius.

\subsection{A uGMRT example}
\label{sec:res_gmrt}

To validate the pipeline with a different instrument, we also ran
CHARIZARD on a uGMRT band-4 observation of the field of the galaxy cluster
PSZ2~G071.21+28.86 \citep{planck2016}. The observation is summarised in
Table~\ref{tab:gmrtobs}. 3C48 is the flux density and bandpass calibrator.
1845+401, 3C468.1 and 3C345 were all identified as phase calibrators. The feeds
are circular and no polarization calibrators were observed, so the pipeline
ran in total intensity mode and kept only $RR$ and $LL$. We
used four sub-bands. Self-calibration ran five phase-only rounds followed by
two amplitude and phase rounds.

\begin{table}
\centering
\caption{The uGMRT observation.}
\label{tab:gmrtobs}
\begin{tabular}{ll}
\hline
\hline
 & PSZ2 G071.21+28.86 \\
\hline
Date & 2024 Mar 11 \\
Band, correlator & Band 4, GWB \\
Frequency (MHz) & $550$--$750$ \\
Channels & $4096 \times 48.8$~kHz \\
Antennas & 30 \\
Reference antenna & C06 \\
Target field & G71+28 \\
Time on target (h) & $3.1$ \\
Flux density, bandpass & 3C48 \\
Phase & 1845+401, 3C468.1, \\
 & 3C345 \\
Polarization & none \\
Image size, cell & $9600^{2}$, $1\farcs0$ \\
\hline
\end{tabular}
\end{table}

The four sub-bands were imaged together over $180$~MHz, giving a
$9600^{2}$ pixel image at $1\arcsec$ per pixel with a
$6\farcs24 \times 4\farcs84$ restoring beam at $650$~MHz. The
\textsc{WSClean} settings are those of Section~\ref{sec:imaging}. Figure~\ref{fig:resgmrt}
shows the inner $1.4^{\circ}$ and the central $16$ arcminutes. The noise is $14.0\,\mu$Jy~beam$^{-1}$. The image is direction independent, and the brightest sources in the field still carry artefacts around them.

We corrected the primary beam with \textsc{Finalflash}\footnote{\url{https://github.com/arpan-52/Finalflash}}. The field lies at a declination of $+44\fdg7$, which is
outside RACS-low. For the positions we use instead the second data release
of the LOFAR Two-metre Sky Survey \citep[LoTSS-DR2;][]{shimwell2022} at
$144$~MHz, whose $6\arcsec$ resolution is very close to ours. For the flux
densities we use NVSS \citep{condon1998} at $1.4$~GHz. The
matching is done the same way as before. \textsc{PyBDSF} at a $5\sigma$
island and pixel threshold returns $555$ sources.

Of the $555$ sources, $361$ have a LoTSS counterpart within $10\arcsec$.
The median offset is $-0.18\arcsec$ in right ascension and $+0.39\arcsec$
in declination, with a median absolute deviation of $0.42\arcsec$ and
$0.37\arcsec$, again under a tenth of the beam (Figure~\ref{fig:resgmrtastro}).

At $\alpha = -0.8$ the median ratio of measured to extrapolated flux
density over the $43$ NVSS sources is $1.13$. If we instead take the
spectral index of each source from LoTSS and NVSS together and interpolate
to $650$~MHz, the median ratio is $1.22$.

\subsection{Performance}
\label{sec:res_perf}

\begin{figure*}
\centering
\includegraphics[width=\textwidth]{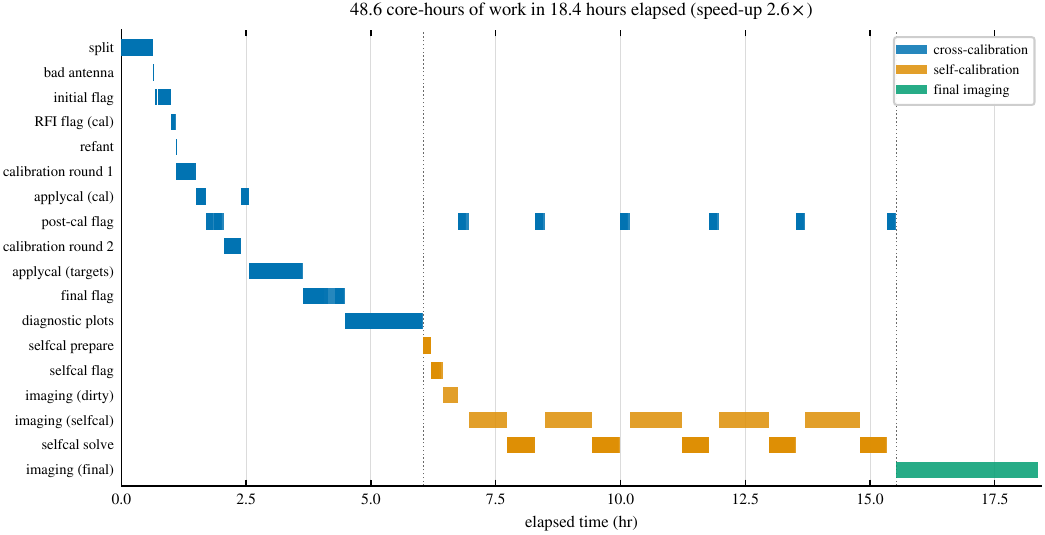}
\caption{Every job of the El Gordo run on a wall-clock axis, from the
\textsc{Housekeeper} database. Each bar is one scheduler job; overlapping
bars ran at the same time, so where the bars stop overlapping is where the
workflow serialises. Colour marks the phase and dotted lines the phase
boundaries. The run performed $48.6$ core-hours of work in $18.4$~hours.}
\label{fig:restime}
\end{figure*}

Every stage of a run is a scheduler job. We use the recorded database, which holds the submission, start and
end time of each job and the cores it held.
Figure~\ref{fig:restime} shows elapsed wall-time for all the jobs from the El Gordo run.

The run completed $48.6$ core-hours in $18.4$~hours of elapsed time. Cross-calibration took
$28.7$ core-hours over $92$ jobs, self-calibration $17.0$ over $34$, and the
final imaging $2.8$ over $4$, one per Stokes parameter.

The shape of Figure~\ref{fig:restime} shows where the parallelism comes from
and where it stops. Through cross-calibration the four sub-bands are
independent, so each stage is four jobs submitted together and costs about
what one sub-band costs. The overlap is wider than that alone would give
because the target fields are flagged concurrently with the calibrator
solves. The self-calibration loop cannot be arranged that way, since each
round must finish before the next begins and the solve waits on the imaging
within a round, so its $17.0$ core-hours spread across $9.3$~hours at a
concurrency below two. Final imaging is four jobs run one after another. In the case of multiple fields, \textsc{CHARIZARD} treats them independently in the self-calibration.

Added up by kind, flagging is $9.9$ core-hours, a fifth
of the run: $1.3$ before calibration, $0.3$ on the calibrators after the
first round, $4.9$ in the post-calibration pass, $2.6$ in the final pass and
$0.7$ inside the self-calibration loop.

\textsc{CHARIZARD} also implements an experimental peeling routine. That would be of interest in the next version of the pipeline release.

\section{Conclusions}
\label{sec:conclusions}

We have presented CHARIZARD, a pipeline that carries radio interferometric
data from a raw measurement set through flagging, cross-calibration,
self-calibration and imaging to full-Stokes images, and which is organised
around distributing that work. The whole of it is expressed as ordinary PBS or SLURM
jobs through \textsc{Housekeeper}, and every computation runs inside a single
rootless container \citep{udocker2018}. The conclusions are as follows:

\begin{itemize}

\item \textsc{CHARIZARD} splits the frequency axis into sub-bands and the
data into independent target fields, and runs each piece as its own
scheduler job. Sub-bands share no bandpass, delay, gain or leakage
solution, and separate targets share no self-calibration solution. The unit of work is the
sub-band through cross-calibration and the field and sub-band pair once
imaging begins.

\item The pipeline calibrates full polarization automatically, for both
linear and circular feeds. \textsc{CHARIZARD} respects models and UV ranges for calibration for known calibrators with provision to supply user defined models and calibration choices.

\item At the end of the self-calibration loop the pipeline images all four
Stokes parameters. The final products include self-calibrated full Stokes images and an optional source list. An experimental direction dependent step is also implemented but not extensively tested.

\item On the El Gordo run with MeerKAT this gave $48.6$ core-hours of work in
$18.4$~hours of elapsed time, a factor of $2.6$, across $130$ jobs. The residual on-axis instrumental polarization is $0.13$ per cent
in linear and $0.06$ per cent in circular, and imaging the leakage
calibrator back from the corrected data confirms it at $0.085$ per cent per
channel. The field reaches $4.11\,\mu$Jy~beam$^{-1}$ over $787$~MHz at
$4\farcs25 \times 3\farcs75$ with $\sigma_I/\sigma_V = 1.010$, which places
the image at its thermal floor.

\item Against RACS-low, positions agree to $-0\farcs17$ in right ascension
and $+0\farcs37$ in declination over $113$ sources, smaller than the
systematic offsets RACS itself shows against other surveys, and flux
densities agree to a median ratio of $1.06$.

\item  On a
uGMRT band-4 observation the image reaches $14.0\,\mu$Jy~beam$^{-1}$ over
$180$~MHz at $6.24 ''\times 4.84''$, positions agree with LoTSS-DR2
\citep{shimwell2022} to $-0.18''$ in right ascension and $+0.39''$ in
declination over $361$ sources, and with median flux density ratio of $13$ compared to 
NVSS \citep{condon1998}.

\item \textsc{CHARIZARD} does not yet correct direction-dependent
polarization. Leakage grows from $0.13$ per cent on axis to $1.01$ per cent
inside the half-power point and $3.54$ per cent beyond it. \textsc{CHARIZARD} still does not do any direction dependent polarization correction.

\end{itemize}

The new age of radio imaging looks promising, and equally heavy on data.
\textsc{CHARIZARD} attempts to address the processing bottleneck with a
framework built on distributed job scheduling and tracking. With a centrally maintained docker image, the pipeline eases the distribution of the packages and serves as an example framework to process current and upcoming interferometer measurements in full polarization.

\section*{Software availability}
CHARIZARD is publicly available at \url{https://github.com/arpan-52/CHARIZARD},
together with \textsc{Housekeeper}, \textsc{Catboss}, \textsc{Foresight} and
\textsc{Finalflash}. The container image is also published as apal52/charizard-pipeline on Docker Hub \footnote{ \url{https://hub.docker.com/r/apal52/charizard-pipeline}}. The runs reported here used the 20260802 tag.

\section*{Acknowledgements}
We acknowledge funding from the Department of Atomic Energy for the project ``Next Generation Instrumentation for Radio Astronomy'' with ID RTI4017.  We thank the staff of the GMRT for making these observations possible. The GMRT is run by the National Centre for Radio Astrophysics of the Tata Institute of Fundamental Research. The MeerKAT telescope is operated by the South African Radio Astronomy Observatory, which is a facility of the National Research Foundation, an agency of the Department of Science and Innovation. This research has made use of the NASA/IPAC Extragalactic Database (NED), which is operated by the Jet Propulsion Laboratory, California Institute of Technology, under contract with the National Aeronautics and Space Administration. The authors acknowledge the use of Claude (Anthropic, Claude  Sonnet 5) for editorial assistance and improving the clarity and flow of this manuscript. 

\bibliography{charizard_gmrt}{}

@book{thompson2017,
  author = {Thompson, A. R. and Moran, J. M. and Swenson, G. W.},
  title = {Interferometry and Synthesis in Radio Astronomy},
  publisher = {Springer},
  address = {Cham},
  year = {2017},
  edition = {3rd}
}

@inproceedings{mcmullin2007,
  author = {McMullin, J. P. and Waters, B. and Schiebel, D. and Young, W. and Golap, K.},
  title = {{CASA} Architecture and Applications},
  booktitle = {Astronomical Data Analysis Software and Systems XVI},
  series = {ASP Conf. Ser.},
  volume = {376},
  pages = {127},
  year = {2007}
}

@article{casa2022,
  author = {{CASA Team}},
  title = {{CASA}, the {Common Astronomy Software Applications} for Radio Astronomy},
  journal = {PASP},
  volume = {134},
  number = {1041},
  pages = {114501},
  year = {2022}
}

@article{offringa2010,
  author = {Offringa, A. R. and de Bruyn, A. G. and Biehl, M. and Zaroubi, S. and Bernardi, G. and Pandey, V. N.},
  title = {Post-correlation radio frequency interference classification methods},
  journal = {MNRAS},
  volume = {405},
  pages = {155},
  year = {2010}
}

@article{offringa2012,
  author = {Offringa, A. R. and van de Gronde, J. J. and Roerdink, J. B. T. M.},
  title = {A morphological algorithm for improving radio-frequency interference detection},
  journal = {A\&A},
  volume = {539},
  pages = {A95},
  year = {2012}
}

@article{offringa2014,
  author = {Offringa, A. R. and McKinley, B. and Hurley-Walker, N. and others},
  title = {{WSClean}: an implementation of a fast, generic wide-field imager for radio astronomy},
  journal = {MNRAS},
  volume = {444},
  pages = {606},
  year = {2014}
}

@article{offringa2017,
  author = {Offringa, A. R. and Smirnov, O.},
  title = {An optimized algorithm for multiscale wideband deconvolution of radio astronomical images},
  journal = {MNRAS},
  volume = {471},
  pages = {301},
  year = {2017}
}

@inproceedings{collier2021,
  author = {Collier, J. D. and Frank, B. and Sekhar, S. and Taylor, A. R.},
  title = {The {IDIA} {processMeerKAT} pipeline: Fast {CASA} processing on a cloud-based {HPC} cluster},
  booktitle = {2021 XXXIVth General Assembly and Scientific Symposium of the International Union of Radio Science (URSI GASS)},
  pages = {1},
  year = {2021},
  doi = {10.23919/URSIGASS51995.2021.9560276}
}

@inproceedings{jozsa2020,
  author = {J{\'o}zsa, G. I. G. and White, S. V. and Thorat, K. and others},
  title = {{MeerKATHI} -- an End-to-End Data Reduction Pipeline for {MeerKAT} and Other Radio Telescopes},
  booktitle = {Astronomical Data Analysis Software and Systems XXIX},
  series = {ASP Conf. Ser.},
  volume = {527},
  pages = {635},
  year = {2020}
}

@inproceedings{kent2020,
  author = {Kent, B. R. and Masters, J. S. and Chandler, C. J. and others},
  title = {Pipeline Calibration and Imaging for the {Very Large Array}},
  booktitle = {Astronomical Data Analysis Software and Systems XXIX},
  series = {ASP Conf. Ser.},
  volume = {527},
  pages = {571},
  year = {2020}
}

@misc{moldon2021,
  author = {Mold{\'o}n, J.},
  title = {{eMCP}: e-{MERLIN} {CASA} pipeline},
  howpublished = {Astrophysics Source Code Library, ascl:2109.006},
  year = {2021}
}

@article{hamaker1996,
  author = {Hamaker, J. P. and Bregman, J. D. and Sault, R. J.},
  title = {Understanding radio polarimetry. {I}. Mathematical foundations},
  journal = {A\&AS},
  volume = {117},
  pages = {137},
  year = {1996}
}

@article{smirnov2011,
  author = {Smirnov, O. M.},
  title = {Revisiting the radio interferometer measurement equation. {I}. A full-sky {Jones} formalism},
  journal = {A\&A},
  volume = {527},
  pages = {A106},
  year = {2011}
}

@article{kenyon2018,
  author = {Kenyon, J. S. and Smirnov, O. M. and Grobler, T. L. and Perkins, S. J.},
  title = {{CubiCal} -- fast radio interferometric calibration suite exploiting complex optimization},
  journal = {MNRAS},
  volume = {478},
  pages = {2399},
  year = {2018}
}

@article{kenyon2025,
  author  = {Kenyon, J. S. and Perkins, S. J. and Bester, H. L. and
             Smirnov, O. M. and Russeeawon, C. and Hugo, B. V.},
  title   = {Africanus {II}. {QuartiCal}: Calibrating radio interferometer data
             at scale using {Numba} and {Dask}},
  journal = {Astronomy and Computing},
  volume  = {52},
  pages   = {100962},
  year    = {2025},
  doi     = {10.1016/j.ascom.2025.100962}
}

@article{perley2013,
  author = {Perley, R. A. and Butler, B. J.},
  title = {An Accurate Flux Density Scale from 1 to 50 {GHz}},
  journal = {ApJS},
  volume = {204},
  pages = {19},
  year = {2013}
}

@misc{mohan2015,
  author = {Mohan, N. and Rafferty, D.},
  title = {{PyBDSF}: Python Blob Detection and Source Finder},
  howpublished = {Astrophysics Source Code Library, ascl:1502.007},
  year = {2015}
}

@phdthesis{briggs1995,
  author = {Briggs, D. S.},
  title = {High Fidelity Deconvolution of Moderately Resolved Sources},
  school = {New Mexico Institute of Mining and Technology},
  year = {1995}
}

@article{cornwell2008,
  author = {Cornwell, T. J.},
  title = {Multiscale {CLEAN} Deconvolution of Radio Synthesis Images},
  journal = {IEEE Journal of Selected Topics in Signal Processing},
  volume = {2},
  pages = {793},
  year = {2008}
}

@article{tasse2018,
  author = {Tasse, C. and Hugo, B. and Mirmont, M. and Smirnov, O. and Atemkeng, M. and others},
  title = {Faceting for direction-dependent spectral deconvolution},
  journal = {A\&A},
  volume = {611},
  pages = {A87},
  year = {2018}
}

@article{vanharlem2013,
  author = {van Haarlem, M. P. and Wise, M. W. and Gunst, A. W. and others},
  title = {{LOFAR}: The {LOw-Frequency ARray}},
  journal = {A\&A},
  volume = {556},
  pages = {A2},
  year = {2013}
}

@inproceedings{jonas2016,
  author = {Jonas, J. and {MeerKAT Team}},
  title = {The {MeerKAT} Radio Telescope},
  booktitle = {Proceedings of MeerKAT Science: On the Pathway to the SKA},
  series = {PoS(MeerKAT2016)},
  pages = {001},
  year = {2016}
}

@article{gupta2017,
  author = {Gupta, Y. and Ajithkumar, B. and Kale, H. S. and others},
  title = {The upgraded {GMRT}: opening new windows on the radio Universe},
  journal = {Current Science},
  volume = {113},
  pages = {707},
  year = {2017}
}

@article{dewdney2009,
  author = {Dewdney, P. E. and Hall, P. J. and Schilizzi, R. T. and Lazio, T. J. L. W.},
  title = {The {Square Kilometre Array}},
  journal = {Proceedings of the IEEE},
  volume = {97},
  pages = {1482},
  year = {2009}
}

@inproceedings{murphy2018,
  author = {Murphy, E. J. and Bolatto, A. and Chatterjee, S. and others},
  title = {The {ngVLA} Science Case and Associated Science Requirements},
  booktitle = {Science with a Next-Generation Very Large Array},
  series = {ASP Conf. Ser.},
  volume = {517},
  pages = {3},
  year = {2018}
}

@article{udocker2018,
  author = {Gomes, J. and Bagnaschi, E. and Campos, I. and David, M. and Alves, L. and Martins, J. and Pina, J. and L{\'o}pez-Garc{\'i}a, A. and Orviz, P.},
  title = {Enabling rootless {Linux} containers in multi-user environments: the {udocker} tool},
  journal = {Computer Physics Communications},
  volume = {232},
  pages = {84},
  year = {2018}
}

@article{sault1996,
  author = {Sault, R. J. and Hamaker, J. P. and Bregman, J. D.},
  title = {Understanding radio polarimetry. {II}. Instrumental calibration of an interferometer array},
  journal = {A\&AS},
  volume = {117},
  pages = {149},
  year = {1996}
}

@article{smirnov2015,
  author = {Smirnov, O. M. and Tasse, C.},
  title = {Radio interferometric gain calibration as a complex optimization problem},
  journal = {MNRAS},
  volume = {449},
  pages = {2668},
  year = {2015}
}

@article{intema2017,
  author = {Intema, H. T. and Jagannathan, P. and Mooley, K. P. and Frail, D. A.},
  title = {The {GMRT} 150 {MHz} all-sky radio survey. First alternative data release {TGSS ADR1}},
  journal = {A\&A},
  volume = {598},
  pages = {A78},
  year = {2017}
}

@article{kale2021,
  author = {Kale, R. and Ishwara-Chandra, C. H.},
  title = {{CAPTURE}: a continuum imaging pipeline for the {uGMRT}},
  journal = {Experimental Astronomy},
  volume = {51},
  pages = {95},
  year = {2021}
}

@article{condon1998,
  author = {Condon, J. J. and Cotton, W. D. and Greisen, E. W. and Yin, Q. F. and Perley, R. A. and Taylor, G. B. and Broderick, J. J.},
  title = {The {NRAO VLA} Sky Survey},
  journal = {AJ},
  volume = {115},
  pages = {1693},
  year = {1998}
}

@article{vandertol2018,
  author = {van der Tol, S. and Veenboer, B. and Offringa, A. R.},
  title = {Image {Domain Gridding}: a fast method for convolutional resampling of visibilities},
  journal = {A\&A},
  volume = {616},
  pages = {A27},
  year = {2018}
}

@article{ranchod2025,
  author  = {Ranchod, S. and Wagenveld, J. D. and Kl\"ockner, H.-R. and
             Wucknitz, O. and Deane, R. P. and Sridhar, S. S. and Barr, E. and
             Buchner, S. and Camilo, F. and Damas-Segovia, A. and Kasemann, C. and
             Kramer, M. and Legodi, L. S. and Mao, S. A. and Menten, K. and
             Rammala, I. and Rugel, M. R. and Wieching, G.},
  title   = {A first glimpse at the {MeerKAT} {DEEP2} field at {S}-band},
  journal = {MNRAS},
  volume  = {536},
  number  = {4},
  pages   = {3647--3662},
  year    = {2025},
  doi     = {10.1093/mnras/stae2754}
}

@article{devilliers2023,
  author  = {de Villiers, M. S.},
  title   = {{MeerKAT} Holography Measurements in the {UHF}, {L}, and {S}
             Bands},
  journal = {AJ},
  volume  = {165},
  pages   = {78},
  year    = {2023},
  doi     = {10.3847/1538-3881/acabc3}
}

@article{kramer2016,
  author  = {Kramer, M. and others},
  title   = {The {MeerKAT} {Max-Planck} {S}-band System},
  journal = {Proceedings of Science},
  volume  = {MeerKAT2016},
  pages   = {003},
  year    = {2016}
}

@article{hale2021,
  author  = {Hale, C. L. and McConnell, D. and Thomson, A. J. M. and
             Lenc, E. and Heald, G. H. and Hotan, A. W. and Leung, J. K. and
             Moss, V. A. and Murphy, T. and Pritchard, J. and Sadler, E. M. and
             Stewart, A. J. and Whiting, M. T.},
  title   = {The Rapid {ASKAP} Continuum Survey Paper {II}: First {S}tokes {I}
             Source Catalogue Data Release},
  journal = {PASA},
  volume  = {38},
  pages   = {e058},
  year    = {2021},
  doi     = {10.1017/pasa.2021.47}
}

@article{mcconnell2020,
  author  = {McConnell, D. and Hale, C. L. and Lenc, E. and Banfield, J. K. and
             Heald, G. and Hotan, A. W. and Leung, J. K. and Moss, V. A. and
             Murphy, T. and O'Brien, A. and Pritchard, J. and Raja, W. and
             Sadler, E. M. and Stewart, A. and Thomson, A. J. M. and
             Whiting, M. and others},
  title   = {The Rapid {ASKAP} Continuum Survey {I}: Design and first results},
  journal = {PASA},
  volume  = {37},
  pages   = {e048},
  year    = {2020},
  doi     = {10.1017/pasa.2020.41}
}

@article{menanteau2012,
  author  = {Menanteau, F. and Hughes, J. P. and Sif{\'o}n, C. and Hilton, M.
             and Gonz{\'a}lez, J. and Infante, L. and Barrientos, L. F. and
             Baker, A. J. and Bond, J. R. and Das, S. and others},
  title   = {The {Atacama Cosmology Telescope}: {ACT-CL} {J0102}$-$4915
             ``{El Gordo}'', a Massive Merging Cluster at Redshift 0.87},
  journal = {ApJ},
  volume  = {748},
  pages   = {7},
  year    = {2012},
  doi     = {10.1088/0004-637X/748/1/7}
}

@inproceedings{lam2015,
  author    = {Lam, S. K. and Pitrou, A. and Seibert, S.},
  title     = {{Numba}: a {LLVM}-based {Python} {JIT} compiler},
  booktitle = {Proc. Second Workshop on the LLVM Compiler Infrastructure in HPC},
  pages     = {1--6},
  year      = {2015},
  doi       = {10.1145/2833157.2833162}
}

@inproceedings{bhatnagar2025,
  author    = {Bhatnagar, S. and Hsieh, M. and Madsen, F. and Jagannathan, P.},
  title     = {{LibRA}: A Scientific Software Library of Radio Astronomy
               Algorithms},
  booktitle = {2025 United States National Committee of URSI National Radio
               Science Meeting (USNC-URSI NRSM)},
  pages     = {282},
  year      = {2025},
  doi       = {10.23919/USNC-URSINRSM66067.2025.10907007}
}

@article{hotan2021,
  author  = {Hotan, A. W. and Bunton, J. D. and Chippendale, A. P. and
             Whiting, M. and Tuthill, J. and Moss, V. A. and McConnell, D. and
             Amy, S. W. and Huynh, M. T. and Allison, J. R. and others},
  title   = {Australian Square Kilometre Array Pathfinder: {I}. System
             description},
  journal = {PASA},
  volume  = {38},
  pages   = {e009},
  year    = {2021},
  doi     = {10.1017/pasa.2021.1}
}

@article{cornwell1981,
  author  = {Cornwell, T. J. and Wilkinson, P. N.},
  title   = {A new method for making maps with unstable radio interferometers},
  journal = {MNRAS},
  volume  = {196},
  pages   = {1067--1086},
  year    = {1981},
  doi     = {10.1093/mnras/196.4.1067}
}

@article{pearson1984,
  author  = {Pearson, T. J. and Readhead, A. C. S.},
  title   = {Image Formation by Self-Calibration in Radio Astronomy},
  journal = {ARA\&A},
  volume  = {22},
  pages   = {97--130},
  year    = {1984},
  doi     = {10.1146/annurev.aa.22.090184.000525}
}

@article{hogbom1974,
  author  = {H{\"o}gbom, J. A.},
  title   = {Aperture Synthesis with a Non-Regular Distribution of
             Interferometer Baselines},
  journal = {A\&AS},
  volume  = {15},
  pages   = {417--426},
  year    = {1974},
  note    = {Predates DOI registration; ADS bibcode 1974A\&AS...15..417H}
}

@article{perley2017,
  author  = {Perley, R. A. and Butler, B. J.},
  title   = {An Accurate Flux Density Scale from 50 {MHz} to 50 {GHz}},
  journal = {ApJS},
  volume  = {230},
  pages   = {7},
  year    = {2017},
  doi     = {10.3847/1538-4365/aa6df9}
}

@article{shimwell2022,
  author = {{Shimwell}, T.~W. and {Hardcastle}, M.~J. and {Tasse}, C. and others},
  title = {{The LOFAR Two-metre Sky Survey. V. Second data release}},
  journal = {A\&A},
  year = 2022,
  volume = {659},
  pages = {A1},
  doi = {10.1051/0004-6361/202142484}
}

@article{planck2016,
  author = {{Planck Collaboration} and {Ade}, P.~A.~R. and {Aghanim}, N. and others},
  title = {{Planck 2015 results. XXVII. The second Planck catalogue of Sunyaev-Zeldovich sources}},
  journal = {A\&A},
  year = 2016,
  volume = {594},
  pages = {A27},
  doi = {10.1051/0004-6361/201525823}
}
\bibliographystyle{aasjournalv7}



\end{document}